\documentclass{webofc}

\usepackage[varg]{txfonts}   
\usepackage{hyperref}
\usepackage{url}
\usepackage{definitions}
\hypersetup{colorlinks=true,citecolor=blue,urlcolor=blue,linkcolor=blue}
\begin{document}
\title{The \nas experiment at SPS}
%
%


\author{
\firstname{Alexander }\lastname{Milov}
\inst{1}\fnsep\thanks{\email{alexander.milov@weizmann.ac.il}} \firstname{} \lastname{for \nas Collaboration}
}

\institute{Department of Particle Physics and Astrophysics, Weizmann Institute of Science, 234 Herzl Street, Rehovot 7610001 Israel}

\abstract{
A new apparatus, \nas, is proposed for measuring muon pairs in the center-of-mass region from 5 to 17 GeV at CERN SPS in various collisional systems from Pb+Pb and down to $p$+Be. The physics scope of the new detector will cover topics from the measurement of thermal radiation coming from the hot and dense medium to chiral symmetry restoration, strangeness, and charm production.

The proposed detector consists of a vertex spectrometer based on novel technology, allowing the production of large silicon sensors and a large-acceptance muon spectrometer based on gaseous detectors. With its high beam intensity, the new apparatus provides access to rare observables that have been scarcely studied until now. The new detector will come into operation after the Long Shutdown 3 of the LHC (past 2029) and is aimed at the first data-taking with Pb and proton beams. In this contribution, we review the project and recent R\&D effort, including the technical aspects and the studies of the physics performances for the observables.
}

\maketitle

\section{Introduction}
\label{sec:intro}
Quark-gluon plasma (QGP) is a state of matter where quarks and gluons are not bound into baryons and mesons but are deconfined over much larger length scales than the hadron size. Such matter can be formed in collisions of Heavy Ions (HI) at high energies, allowing us to scrutinize our Quantum Chromodynamics (QCD) knowledge in laboratory experiments using colliders and beams extracted from accelerators and dumped onto fixed targets.

The QGP created in the interactions of heavy ions and studied by the experiments is characterized by the temperature $T$ and baryo-chemical potential $\mu_{\rm{B}}$. High-energy colliders such as LHC and RHIC produce matter at nearly zero net baryonic density (or equivalently $\mu_{\rm{B}}=0$) and temperature $T$ reaching about 500 MeV. Collisions in the center-of-mass energy range per nucleon-nucleon collision in the range $6 < \sqn < 17.3$ GeV, as available at the CERN SPS, may lead to the formation of a QGP, characterized by a smaller initial $T$ and non-zero $\mu_{\rm{B}}$. When increasing $\mu_{\rm{B}}$, the transition from hadronic matter to QGP is expected to change from a crossover to a first-order phase transition, with a critical point separating the two regimes.

\begin{figure}[h]
\centering
\includegraphics[width=0.5\textwidth]{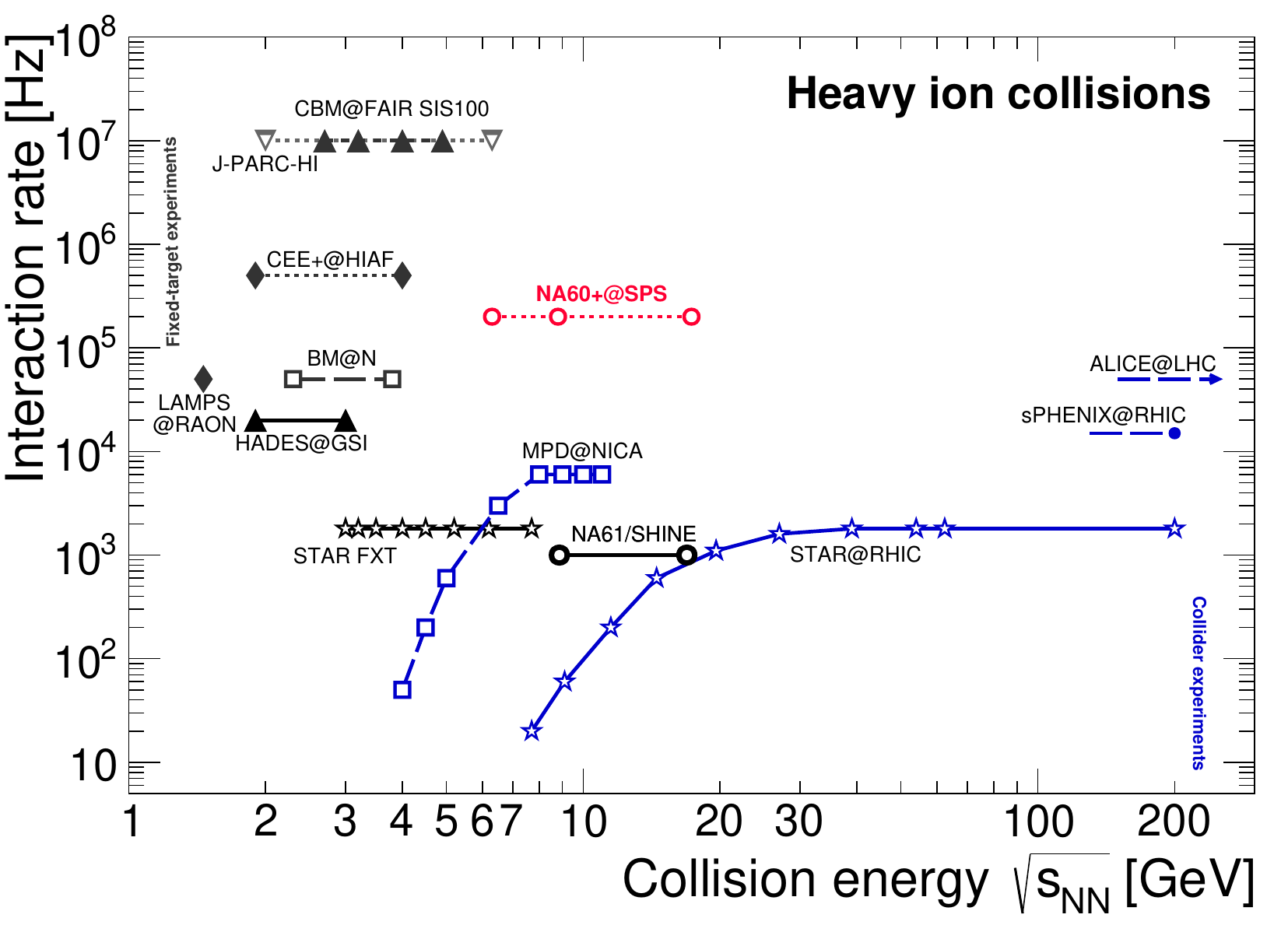}
\caption{Operational and planned HI experiments. Plot is taken from Ref.~\cite{Galatyuk:2019lcf}.}
\label{fig:rates}       
\end{figure}

\begin{figure*}[h!]
\centering
\includegraphics[width=0.75\textwidth]{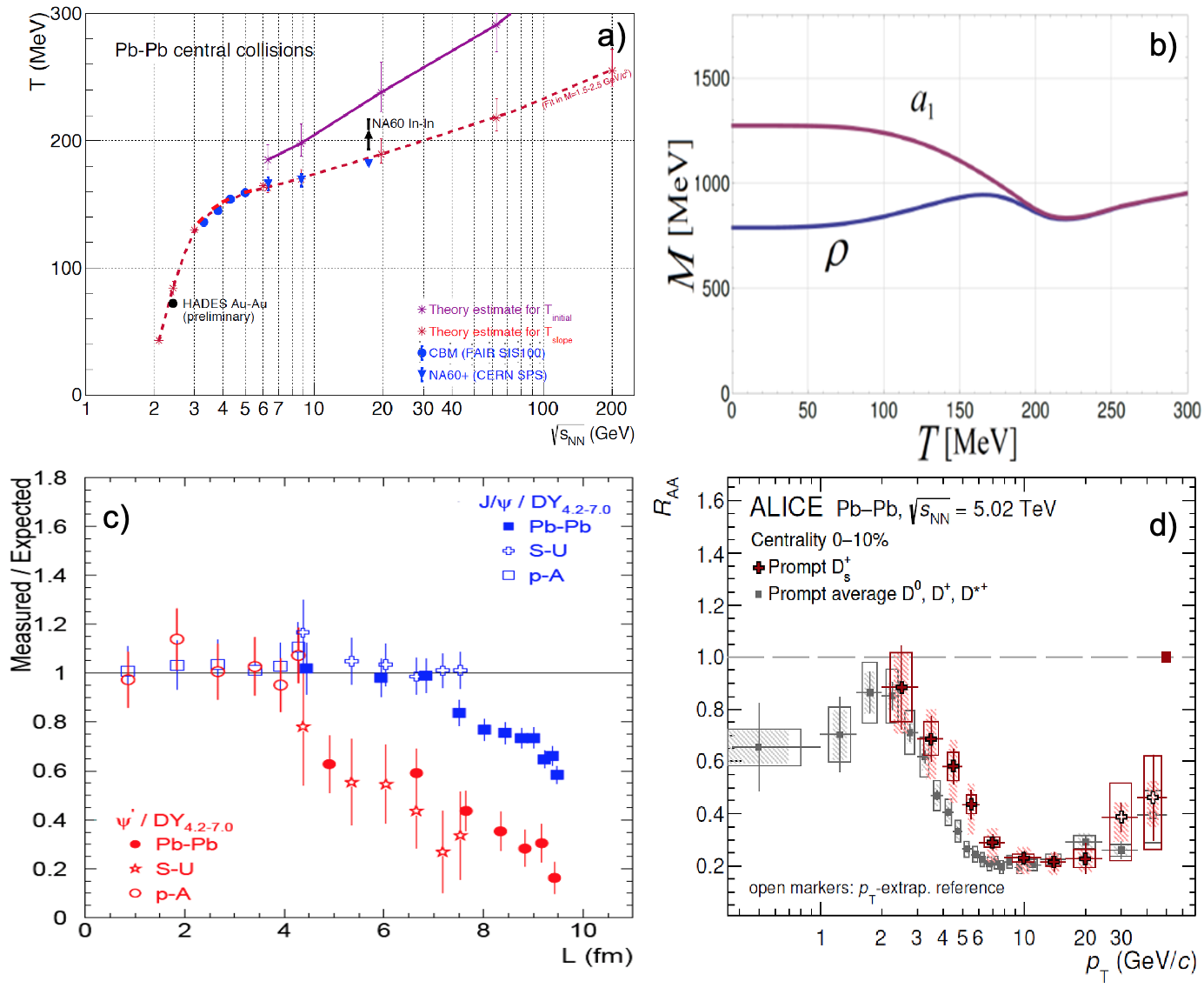}
\caption{Observables to be measured by the \nas experiment at the SPS. Panel a) is the caloric curve, showing projected results of the \nas. Panel b) is the $\rho$- and $a_{1}$-meson mixing. Panel c) is the \jpsi over Drell-Yan production, and Panel d) is the nuclear modification factor of open charm mesons. The curves are taken from Refs.~\cite{Rapp:2014hha, Galatyuk:2015pkq, Jung:2016yxl, NA50:2000brc, NA50:2006yzz, ALICE:2021kfc}.}
\label{fig:goals}
\end{figure*}

\nas is the new experiment proposed at the CERN SPS~\cite{NA60:2022sze}, focusing on studying specific observables related to the high-$\mu_{\rm{B}}$ QGP formation and the corresponding phase transition. The \nas occupies a unique place among existing and planned facilities, allowing data collection at the highest rate in SPS \sqn interval of energies, see Figure~\ref{fig:rates}.
The \nas will use muons to study electromagnetic probes of the QGP, and measure hidden charm production. Open charm will be measured using detector tracking capabilities. The former gives access to the temperature of the deconfined medium and to the modification of the hadronic spectrum due to the restoration of the chiral symmetry of QCD close to the phase transition. The latter gives constraints on the transport properties of the QGP (open charm) and on the modification of the QCD binding in a deconfined medium (charmonium). None of these observables can be accurately measured in the SPS energy range by any other existing or presently foreseen experimental program. Main observables related to thermal radiation, strangeness, and charm production are summarized in Fig.~\ref{fig:goals}.

\section{The apparatus}
\label{sec:setup}
\begin{figure}[ht]
\begin{center}
\includegraphics[width=0.5\textwidth]{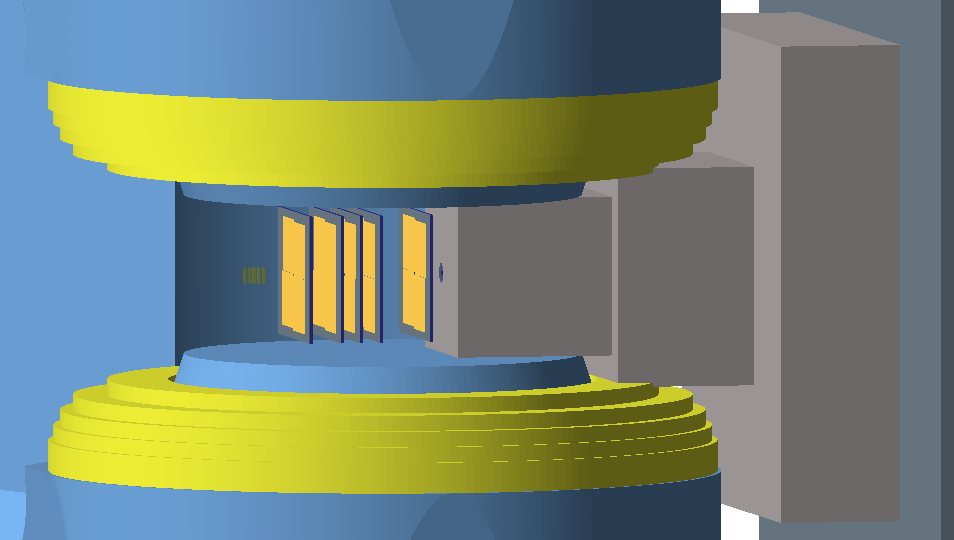} \hspace{5mm}
\includegraphics[width=0.35\textwidth]{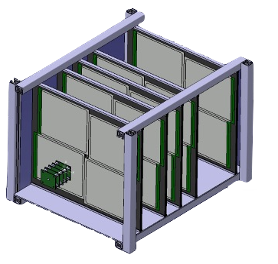}
\caption{Left: \nas VS inside the  MEP48 dipole magnet. Right: Zoom in on the VS with an assembly of 5 targets positioned in the front of the VS.}
\label{fig:detectors:vertex-telescope-fig2}
\end{center}
\end{figure}

The \nas detector consists of two major subsystems. The vertex spectrometer (VS) and the muon system (MS). The former measures the kinematic parameters of muons before they reach the hadron absorber. The VS also reconstructs tracks of charged hadrons, enabling further studies on open charm and strangeness production. The VS consists of 5 silicon pixel stations immersed in the 1.47 T dipole field of MEP48 magnet. The total active area of the silicon sensors in the VS is approximately 0.5 m\(^2\). The VS is shown in Fig.~\ref{fig:detectors:vertex-telescope-fig2}.
The VS readout sensors, 4 in each station, are developed in collaboration with ALICE ITS3. The sensors use 25~mm long units, replicated and stitched together. Stitched sensors having $15\times15$ cm$^2$ in transverse dimensions have a thickness of 0.1\% of a radiation length and 5 $\mu$m resolution. Sensors are mounted on frames, the design of which is being finalized. The frames provide mechanical support and cooling required for operating the sensors. The first large-area sensors are expected to be produced in 2026.

\begin{figure}[ht]
\begin{center}
\includegraphics[width=0.37\textwidth]{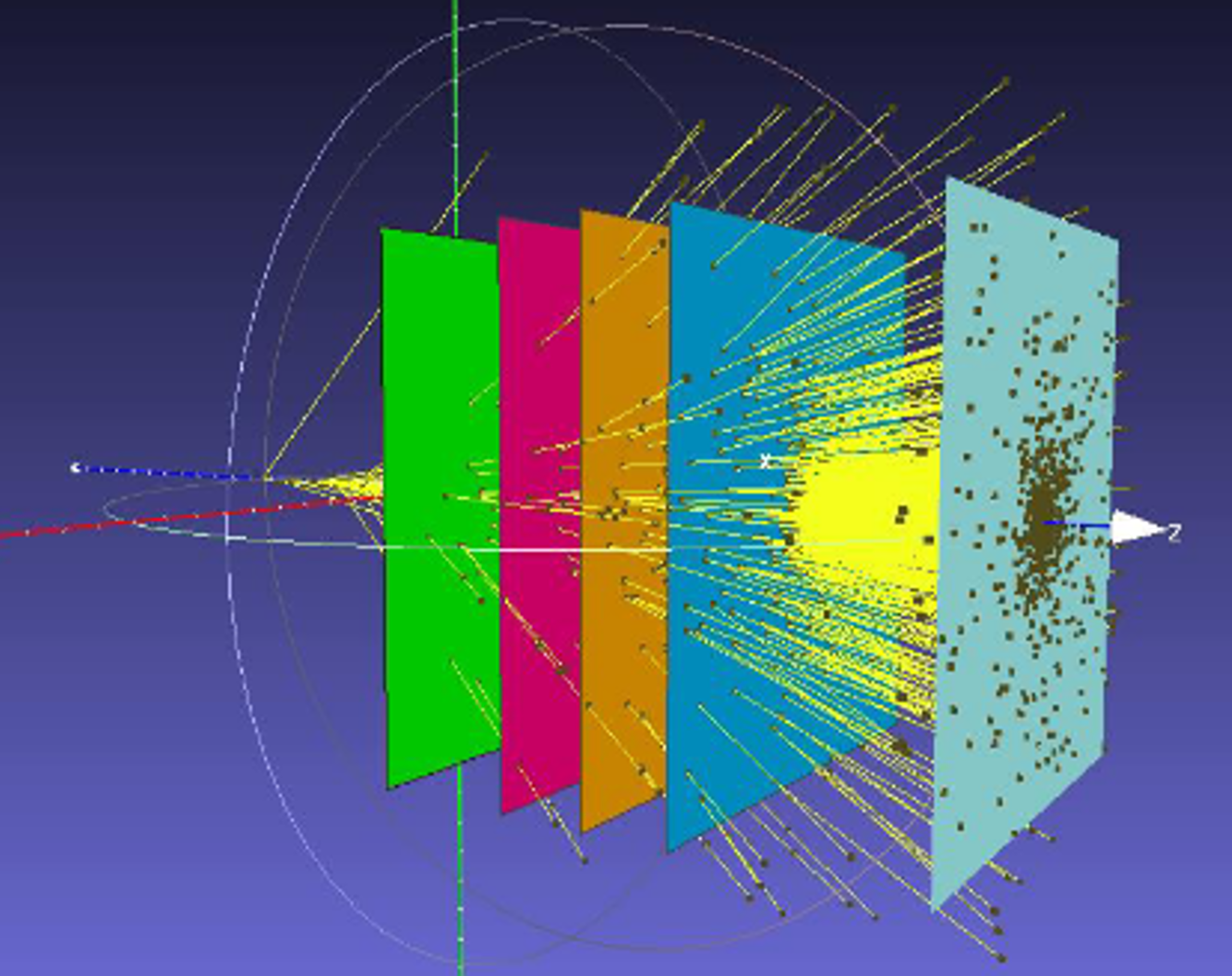}
\includegraphics[width=0.5\textwidth]{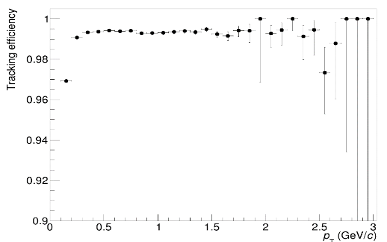}
\caption{Left: Tracks reconstructed in the VS using ATCS package. Right: tracking efficiency.}
\label{fig:tracking}
\end{center}
\end{figure}

Figure~\ref{fig:tracking} shows tracks in the VS reconstructed with the ACTS tracking software package~\cite{acts}.
Tracks reconstructed in the VS spectrometer are shown in the left panel of the figure, and their tracking efficiency, rising above 100 MeV, is shown on the right. These initial results demonstrate that the \nas collaboration has made significant progress on implementing tracking algorithms, which have recently become working also for the MS. 

\begin{figure}[h!]
\begin{center}
\includegraphics[width=0.34\textwidth]{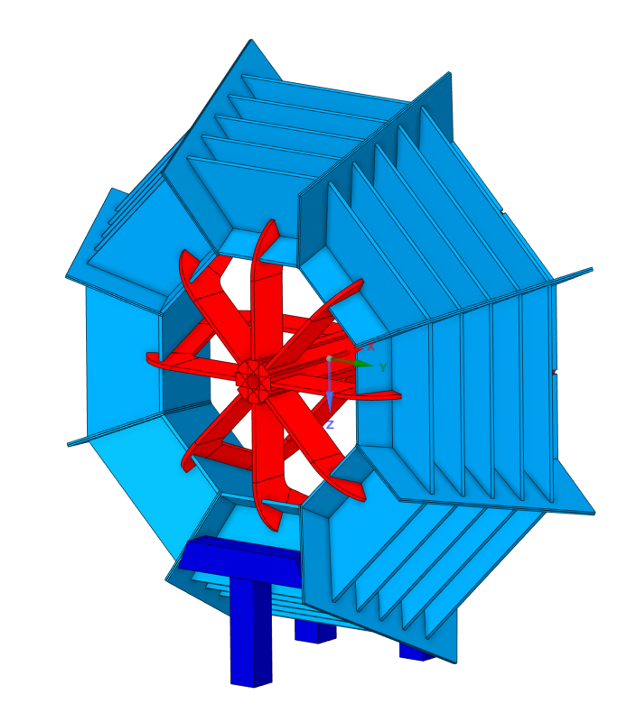}
\includegraphics[width=0.53\textwidth]{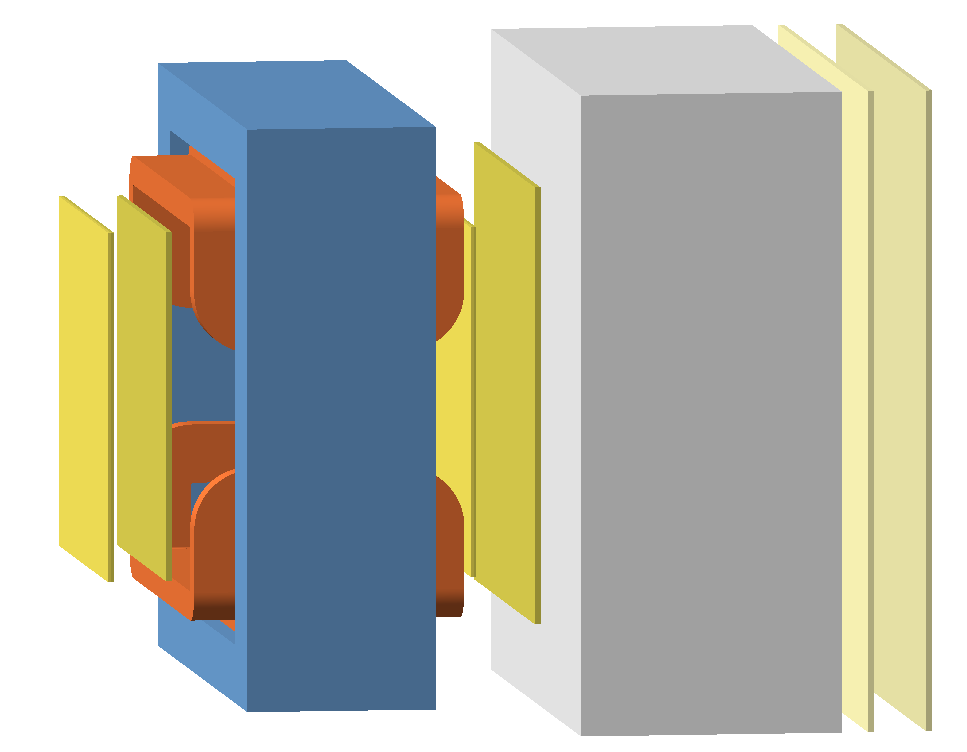}
\caption{Left: The latest design of the toroidal magnet worked out by CERN Magnets, Superconductors and Cryostats group for the \nas experiment. Right: The MS layout based on the MNP33 magnet. The magnet yoke and coils are shown in blue and red, respectively. Readout chambers are shown in yellow, and the additional absorber is shown in grey.}
\label{fig:magnets}
\end{center}
\end{figure}

The primary function of the MS is to measure the kinematic parameters of tracks penetrating the absorber. The MS measures tracks in the magnetic field of the second magnet. Until recently, the \nas collaboration considered building a toroidal magnet in a way similar to the scheme of the NA60 experiment. The magnet design was worked out by the CERN Magnets, Superconductors and Cryostats group as shown in the left panel of Fig.~\ref{fig:magnets}. 
The magnet satisfying the demands of \nas reached almost 8~m in diameter and required significant reinforcements to support the structure and sustain deformations due to the magnetic field forces. The optimization made by the CERN group showed that the construction of the magnet and its integration in the experiment neared the cost of 5 MChF. The size of the magnet would also require civil construction in the experimental hall of the H8 beam line, foreseen as the location of the experiment. 

An alternative solution was found after the decision to end the experimental program of the NA62~\cite{NA62:2017rwk} experiment, measuring rare kaon decays. Currently, the data-taking is ongoing, and the MNP33 magnet~\cite{Fry:2016sjj} will become available before the start of the \nas experimental program in 2029. The main change needed to be introduced in the \nas setup to use the MNP33 magnet is due to the fact that MNP33 is a dipole. The current MS configuration using the dipole is shown in the right panel of Fig.~\ref{fig:magnets}.

MS stations measure positions of charged particle hits before and after the magnet, determining their kinematics and momentum. For muons, the momentum resolution in the MS is dominated by fluctuations in energy loss, which the muons suffer as they pass the absorber. The MS tracks can be matched with those measured in the VS before the absorber and accepted as muon candidates. Matching tracks drastically improves the resolution of muon reconstruction. Based on simulation, to ensure an effective matching, readout chambers should possess a spatial resolution of about 200 $\mu$m. 

Two additional chambers positioned behind the absorber wall function as muon identifiers. Those chambers cover larger areas and have a coarser resolution. The area of all stations covers approximately 75~m$^2$. The MS stations are made of overlapping modules, each less than 1~m in transverse dimension. Their design and final dimensions are being worked out. The technology choice for the MS readout modules is the multi-wire proportional chambers with stripped cathode readout. This technology provides the highest reliability at the lowest cost, while generally satisfying all requirements of the experiment. Alternative technologies based on micro-pattern gaseous detector technology are also considered for the highest hit density areas, predominantly in the first stations. 

\begin{figure}[h!]
\begin{center}
\includegraphics[width=0.30\textwidth]{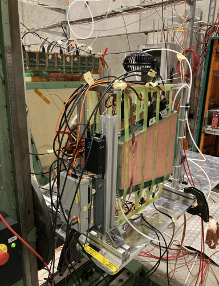} \hspace{3mm}
\includegraphics[width=0.30\textwidth]{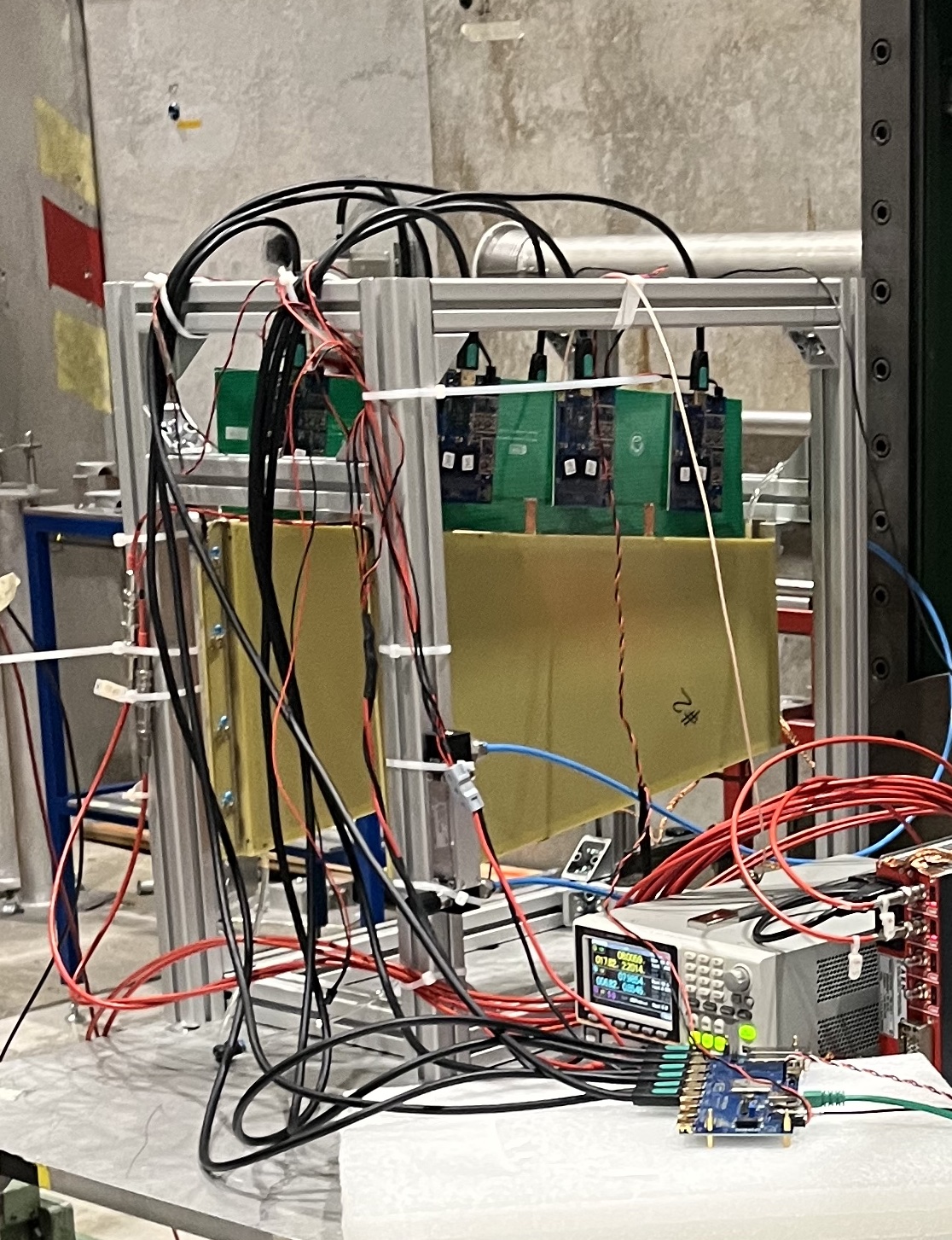}
\caption{The first (left) and the second (right) prototypes of the MS readout modules installed in the hall of the H8 beam line at CERN. The trapezoidal shape was considered for the toroidal magnet configuration. The prototype on the right is equipped with the first version of electronics produced by the USTC group.}
\label{fig:proto}
\end{center}
\end{figure}

The two prototypes of the readout modules shown in Fig.~\ref{fig:proto}.  
were produced by the group of the Weizmann Institute of Science and the group of the University of Science and Technology of China. The prototypes installed in the test beam at CERN were tested in 2023 and 2024 using the SPS Pb beam incident on Pb and other targets installed in the H8 beam line.

\begin{figure}[ht]
\begin{center}
\includegraphics[width=0.27\textwidth]{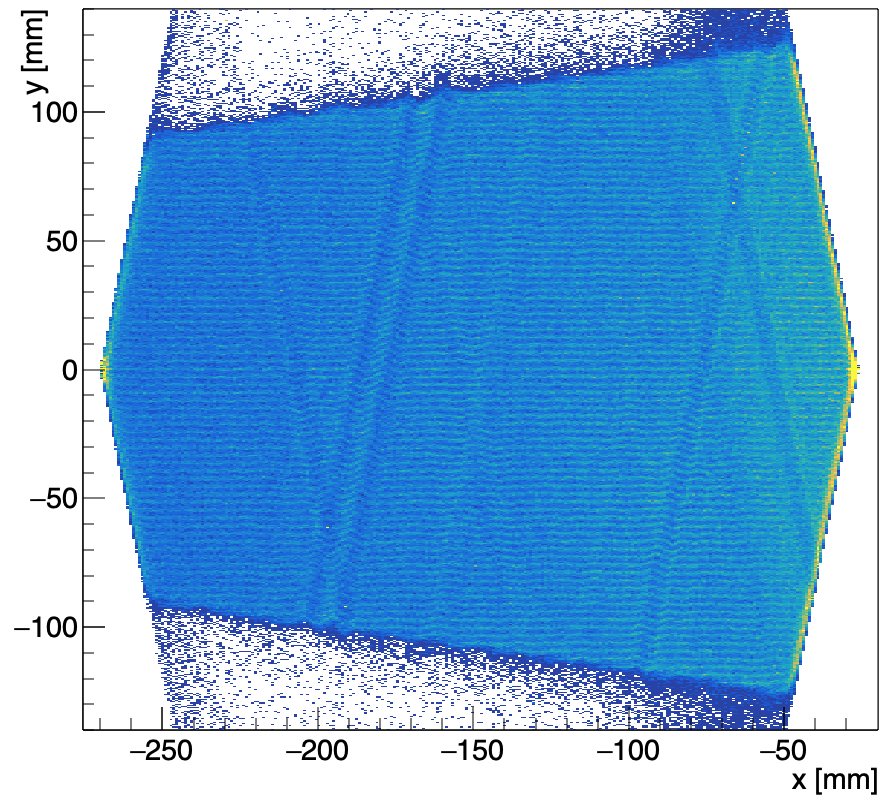}
\includegraphics[width=0.31\textwidth]{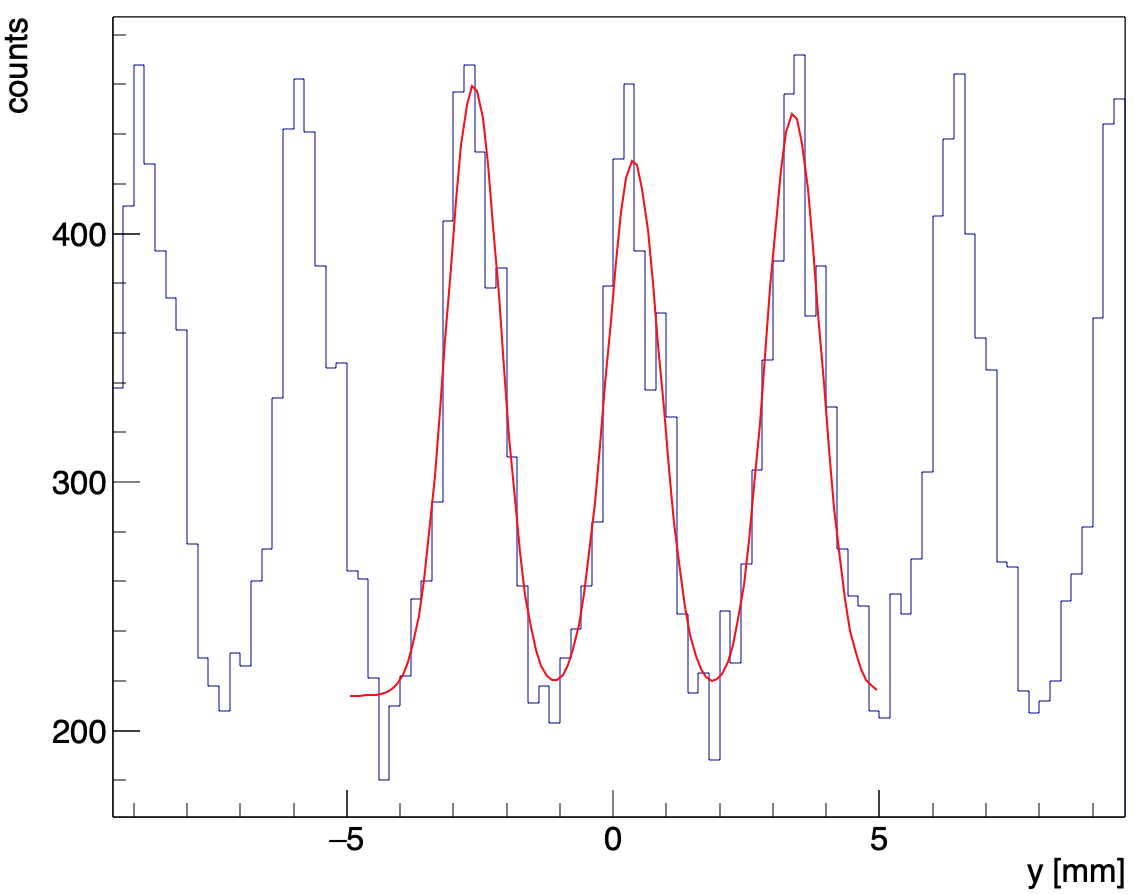}
\includegraphics[width=0.40\textwidth]{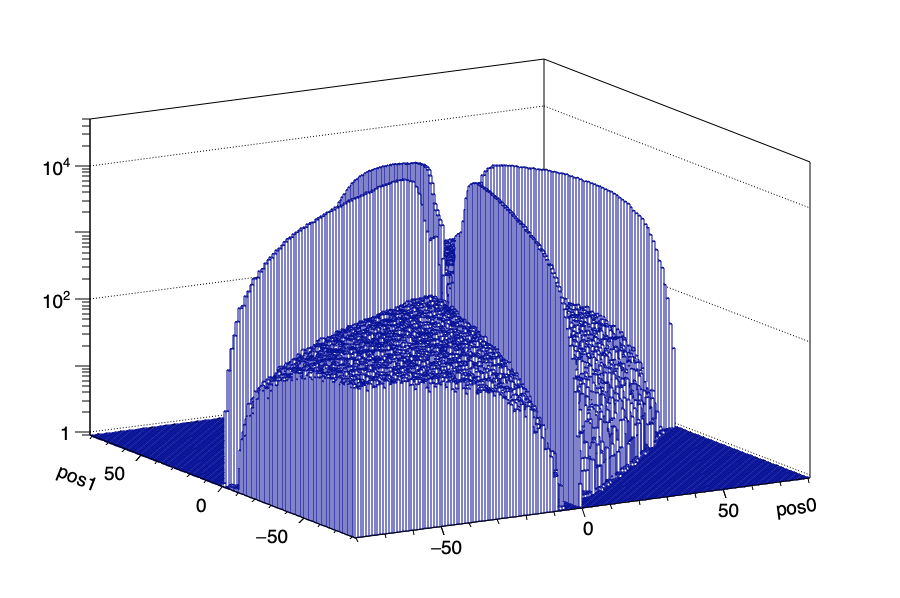}
\caption{Left: A 2D hit map in the MS readout chamber prototype. Middle: wire position measured from strips with peak width fitted to 500~$\mu$m. Right: distance between the hits measured in prototypes. Tall ridges are hits merged in one of the strips.}
\label{fig:test_beam}
\end{center}
\end{figure}

Several results were obtained with the test beams of the prototypes that are shown in Fig.~\ref{fig:test_beam}.
The 2D hit map of the prototype illuminated by secondary particles is shown on the left panel of the Figure. It was demonstrated that the spatial resolution required by the experiment can be achieved. Based on the strip structure used in the prototype, it measured to be 500~$\mu$m across the wires and 100~$\mu$m along the wires. These parameters will change during the detector optimization. Since no absorber was installed in the test beams, the chambers were operational in a much larger multiplicity than they would be in the experiment. That allowed the investigation of the strip structure to resolve double hits. It is shown in the right panel of Fig.~\ref{fig:test_beam} as a 2D histogram of distances between the hits. The tall ridges are expected to correspond to when hits merge in one of the two strip directions. This allows further optimization of the strip structure, which is currently ongoing. Several changes in the design of the prototype were made after the first round of tests, aiming to improve the chamber design.

\section{The plans}
\label{sec:plans}
The \nas collaboration plans to submit the proposal for the new experiment at SPS under the tentative name Dimuon and Charm Experiment (DiCE) in May 2025. The work on the design of the experiment is ongoing to start the construction of apparatus elements in 2026-27 and be ready for the data taking at the end of the Long Shutdown 3.


\vspace{5mm}
The work of the presenter and collaborators is supported by the grant number 714777 from the Israeli Science Foundation and the grant number 12361141827 from the National Natural Science Foundation of China.

%
%
%

\end{document}